\documentclass[prl,twocolumn,showpacs]{revtex4-1} 
\pdfoutput=1
\usepackage{txfonts}
\usepackage{graphics}
\usepackage{graphicx}
\begin{document}

\title{Role of the granular nature of meteoritic projectiles in impact crater morphogenesis}

\author{Roberto Bartali}
\author{Gustavo M. Rodr\'{\i}guez-Li\~{n}\'{a}n}
\author{Yuri Nahmad-Molinari}\email{yuri@ifisica.uaslp.mx}
\author{Damiano Sarocchi}
\affiliation{Universidad Aut\'{o}noma de San Luis Potos\'{\i}, \'{A}lvaro Obreg\'{o}n No. 78, 78000, San Luis Potos\'{\i}, San Luis Potos\'{\i}, Mexico}
\author{J. C. Ruiz Su\'{a}rez}
\affiliation{CINVESTAV-Monterrey, PIIT Autopista Nueva al Aeropuerto Km. 9.5, Apodaca, Nuevo Le\'on 66600, Mexico}
\date{\today}
\pacs{45.70.-n, 96.30.Ys, 96.20.Ka}

\begin{abstract} 
By means of novel volume-diameter aspect ratio diagrams, we ponder on the current conception of crater morphogenesis analyzing crater data from beam explosions, hypervelocity collisions and drop experiments and comparing them with crater data from three moons (the Moon, Callisto, and Ganymede) and from our own experimental results. The distinctive volume-diameter scaling laws we discovered make us to conclude that simple and complex craters in satellites and planets could have been formed by granular-against-granular collisions and that central peaks and domes in complex craters were formed by a dynamic confinement of part of the impacting projectile, rather than by the uplift of the target terrain. A granulometric analysis of asteroids and central peaks and domes inside complex craters, shows boulder size distributions consistent with our hypothesis that crater internal features are the remnants of granular impactors.
\end{abstract}

\maketitle

Although the origin of craters in planetary bodies of the Solar System is no longer an astrophysical puzzle \cite{Pike1974,Wegener1975,Shoemaker1977} and the root of their morphology seems to be a settled issue \cite{Melosh1989,Melosh1999,Holsapple1993}, the evidence that increasingly builds up on the granularity of asteroids \cite{Housen2003,Asphaug2002,Fujiwara2006,Scheeres2010} hints towards a critical revision of the existing paradigm. Succinctly, the paradigm of crater formation, which fails to explain certain features like the existence of boulders perched on central peaks, says that craters and central peaks were created by the target fluidization produced by solid meteorite impacts, followed by collapse of the transient crater where final shapes of craters emerge \cite{Melosh1989,Melosh1999,Holsapple1993}. Analyzing crater data generated in terrestrial experiments \cite{Holsapple2004a} from beam explosions \cite{Holsapple2004b}, hypervelocity collisions \cite{Holsapple2004c} or drop experiments \cite{Holsapple2004d} during the last four decades, could help in understanding crater morphogenesis if we can compare them with observational crater data gathered from different moons \cite{Schenk2002,Losiak2009} at the light of recent \cite{Pacheco-Vazquez2011} experimental results in crater formation.

Impact craters should be considered as the etched scars of the accretion process of planetary formation. Dipole to dipole induced interactions or London dispersive forces and inelastic collapse, where dissipative collisions excite radiative degrees of freedom \cite{Goldhirsch1993}, are the firsts mechanisms of accretion of cosmic dust into small objects hold together by these cohesive forces, that eventually become large granular piles, asteroids and planetesimals, whose growth is driven afterwards by the much smaller gravitational force \cite{Scheeres2010}.

We could describe the first generation of accreted minor bodies as ``astroprotoliths'', which are essentially a pile of granulated rock-forming elements and ices. These softly consolidated objects could very likely represent the vast majority of impactors during the planet formation period and afterwards. The pressure in the cores of astroprotoliths is not enough to trigger or sustain metamorphic processes of lithification until they reach enough mass to initiate lithification and differentiation \cite{Asphaug2002,Scheeres2010} at a critical diameter of 250~km; as can be calculated by solving the hydrostatic gravitational equilibrium equation for the 1500~bar critical pressure \cite{Blatt1996} needed for triggering the metamorphism. A second generation minor bodies can be thought as a set of small solid objects that result from cataclysmic collisional events in which well consolidated planetoids were broken apart into a myriad of fragments. Metallic asteroids that only could be produced within large objects differentiated by geological processes are examples of such class. 

\begin{figure}[tb]
\noindent\resizebox{\columnwidth}{!}{\includegraphics{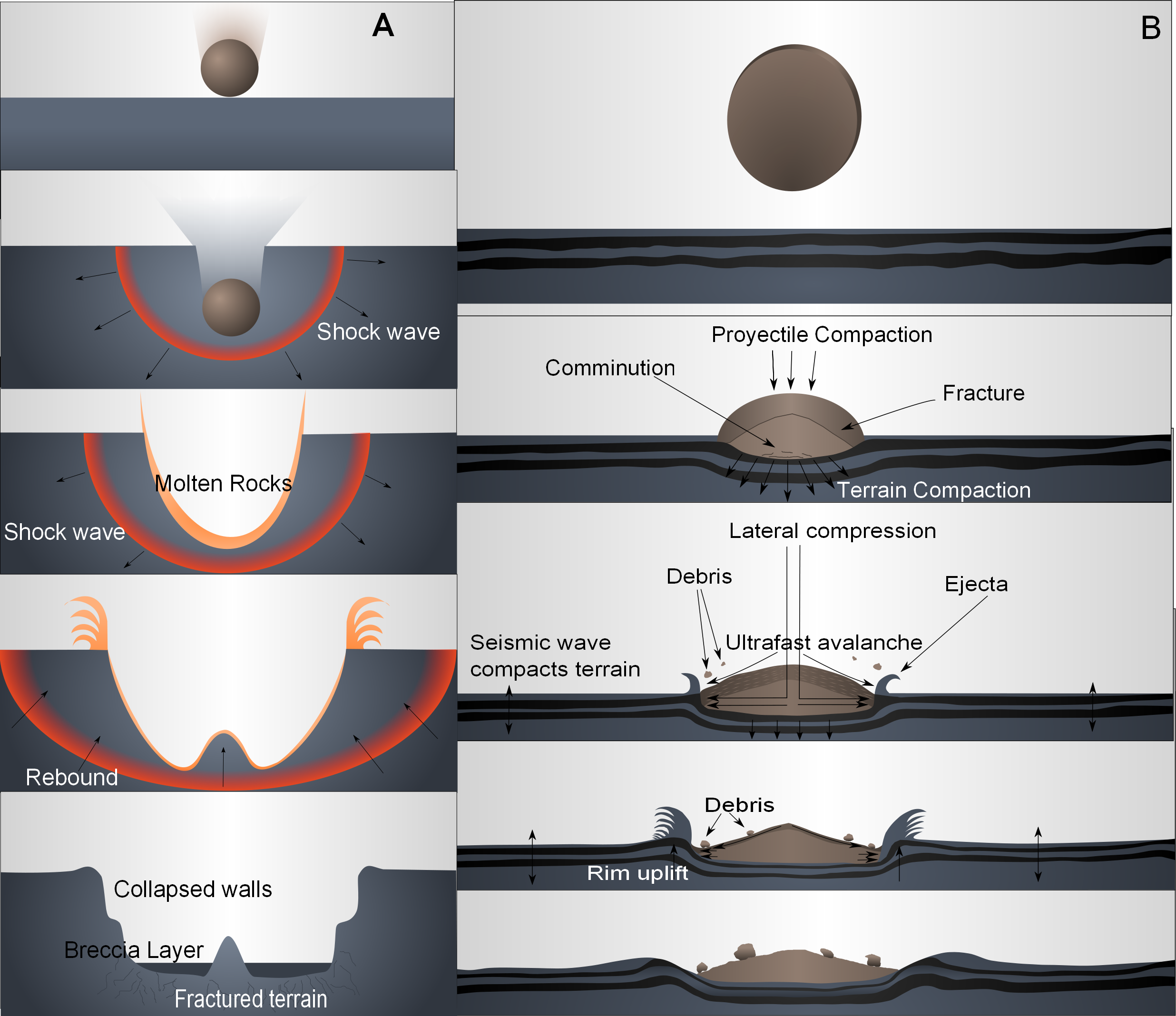}}
\caption{\label{Impact}(A) Solid-against-solid impact, where the compression, excavation, and modification stages are depicted. (B) Granular-against-granular impact, where the comminution and moldering of the projectile is shown. The measured stratigraphy after impact, from which local apparent compaction (area changes) can be inferred.}
\end{figure}

Current opinion in impact cratering phenomena depicts the following scenario: contact and compression of the impactor and the terrain followed by an explosive excavation of the so called transient crater, produced by a sudden melting and evaporation processes of the impactor, the uplift of the decompressed terrain, and the final collapse of the transient crater \cite{Melosh1989,Melosh1999}. This scenario, schematically depicted in Fig.~\ref{Impact}A, and only accounted in its details by computational codes that solve a set of conservative and constitutive equations \cite{Pierazzo2004}, occurs when the impactor is a solid lithified structure and the escape velocity satisfies 
\begin{equation}
v_e>\sqrt{L_f+\Delta TC_v},
\end{equation}
where $L_f$  is the latent heat of fusion of the impactor body and $C_v$ is the specific heat capacity. Moreover, the critical radius, $R_p$ of the target having a mean density $\rho_p$ that allows evaporation of the impactor is of the order of 4~000~km
\begin{equation}
R_p=\sqrt{\frac{L_f+\Delta TC_v}{G\rho_p}}.
\end{equation}

However, for smaller target bodies, first generation minor impactors and gravitationally driven collisions, a very different chain of events may take place. Compression of the astroprotolith does not occur in supersonic collisions, since elastic deformations are unable to propagate along the body before shear stresses around the immediate vicinity of the contact point reach their yield value (which is several orders of magnitude smaller than for well consolidated solid objects). Thus, as a first step a supersonic comminution or moldering of the body is the first event to occur. Simultaneously, the granular flux of the moldered astroprotolith will exert a dynamic pressure on the surface of the target producing a subsidence by plastic deformation of the impacted terrain and just a small fraction of the impactor material penetrates in it, due to huge hydrodynamic drag exerted on such small particles by the compressed terrain. 

A comparison of the penetration of a large solid body with a that of a comminuted body of similar size composed by small particles of a fraction $S$ of the radius of the solid one, can be done by modeling it as a motion of objects through a viscous fluid, and supposing that all the original kinetic energy is dissipated doing work against friction. Considering two particles of equal density and size ratio of $S = R / r$, moving at the same velocity, their kinetic energies would have a ratio of $S^3$, and similarly the frictional force ratio among them scales up as $S^2$, thus, the penetration length of the entire body will be $S$ times larger than the penetration length of their fragments. If we think of a 500~m astroprotolith mainly composed of millimeter sized fine grains of interplanetary dust, their penetration will differ by a factor of 500,000.

\begin{figure}[tb]
\noindent\resizebox{\columnwidth}{!}{\includegraphics{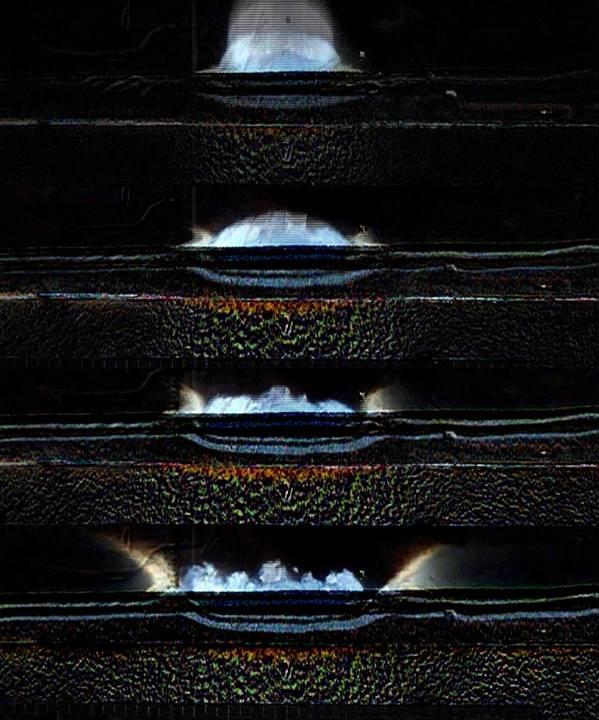}}
\caption{\label{Diff}Two-dimensional experiment of granular-against-granular crater formation as captured by differential images, in which displacements (obtained by subtracting subsequent frames) are represented in false color.}
\end{figure}

In this sense, ballistic penetration of fragile or prone to be comminuted objects is unlikely, and the target terrain will suffer instead a huge dynamic pressure, due to the ultrafast avalanche of the sand pile formed during the very collision event. This ultrafast avalanche compress the terrain while the material which makes first contact with the target in the center of the vertical avalanche would get trapped between the terrain itself and the granular flux of the comminuted body \cite{Amarouchene2001}, conforming the central peak and dome structures that will remain after collision. This mechanism has been called dynamic confinement \cite{Pacheco-Vazquez2011} and turns out to be the most likely way to interpret the origin of peaked and domed craters observed in the Solar System bodies, instead of invoking rebound and uplift of the impacted terrain, not observed yet experimentally. In Fig.~\ref{Diff}, a series of pictures showing the central peak formation in 2D impact experiments reveals the robustness of this crater opening mechanism for granular-granular collisions.

It should be stressed out that the compressional mechanism of crater formation just described, significantly differs from the crater formation due to solid objects impacting on sand, fragile solids or even explosive craters produced by bombs (something we have tried to sketch in the artistic drawings shown in Fig.~\ref{Impact}). The main difference resides in the fact that this kind of craters implies the mobilization of a certain quantity of the targets material which is proportional to the impactor volume itself or the explosive power of the load. In other words, the material displacement of the target proceeds in three dimensions without any restriction in the excavation mechanism. However, for granular-granular impacts, the crater formation mechanism in the vertical direction is mainly a compressional subsidence of the target terrain and it has to be taken into account the fact that the impactor material fills out partially the crater just formed (see Fig.~\ref{Impact}B). Furthermore, compressive forces for granular materials obey a highly non linear equation \cite{Frenning2009} that saturates for displacements that produce volume fractions close to one. These two factors (saturating compressibility and partial filling off the crater by the impactor material), impose a severe constriction to the crater excavation mechanism in the vertical direction and should lead to a nontrivial scaling law of the aspect ratio plots. 

\begin{figure}[tb]
\noindent\resizebox{\columnwidth}{!}{\includegraphics{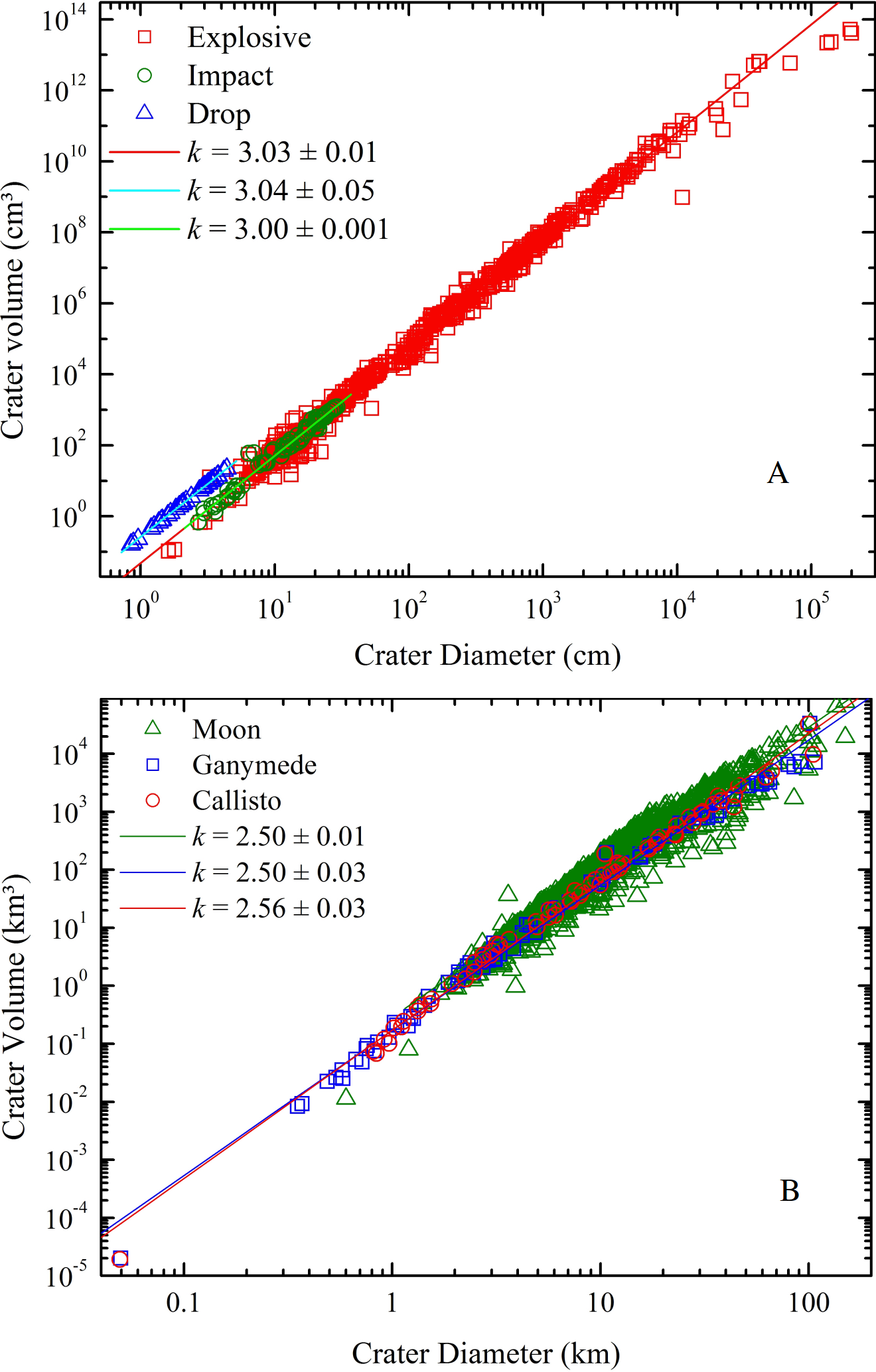}}
\caption{\label{VvsD}(A) Volume-diameter plots of different craters. Dropping of low speed (blue triangles), hypersonic projectiles (green circles), and explosive (red squares) experimental results and their respective fits, showing power laws with exponents very close to 3. (B) Volume-diameter plots for satellite craters of three bodies: the Moon (green triangles), Ganymede (blue squares), and Callisto (red circles) and their corresponding linear fits, showing power laws with exponents close to 2.5.}
\end{figure}

\begin{figure}[tb]
	\noindent\resizebox{\columnwidth}{!}{\includegraphics{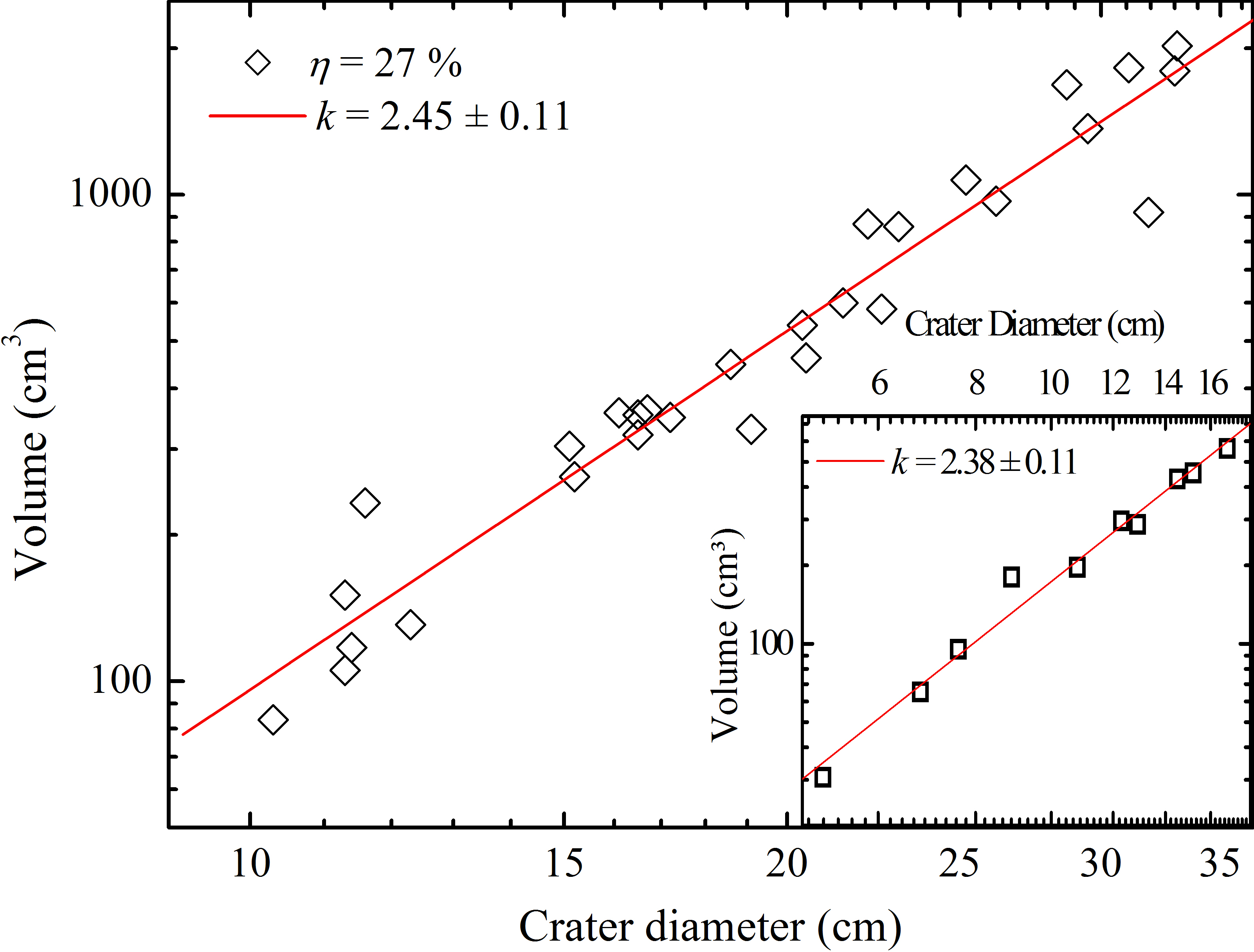}}
	\caption{\label{Exper}Volume-diameter plots for granular-granular impact craters. Data obtained for this article (main figure) and the experiments from Pacheco-V\'azquez and Ruiz-Su\'arez \cite{Pacheco-Vazquez2011} (inset), showing a power law with exponent close to 2.4}
\end{figure}

Data from explosion experiments \cite{Holsapple1993,Holsapple2004b}, shooting solid impactors against granular targets at high speeds \cite{Holsapple2004c}, and dropping solid objects on sand \cite{Holsapple2004d}, plotted in a crater volume versus crater diameter diagram \cite{Cintala1998}, show the crater morphology depicted in Fig.~\ref{VvsD}A. These experiments share in common a power law growth of the volume crater with increasing crater diameter, with exponent very close to 3 (Fig.~\ref{VvsD}) for more than 14 decades in volume. Similarly, more recent experiments of hypervelocity impacts on sandstone target performed by Dufresne \emph{et al.} give rise to a similar exponent if one plots the crater volumes as a function of the crater diameters they report\cite{Dufresne2013}. Surprisingly, if we plot the same kind of aspect ratio plots for craters observed in the Moon \cite{Losiak2009}, Ganymede and Callisto \cite{Schenk2002} (Fig.~\ref{VvsD}B), we observe a power law with exponents between 2.35 and 2.5, respectively. For comparison, we plot in Fig.~\ref{Exper} experimental data from Pacheco-V\'azquez and Ruiz-Su\'arez \cite{Pacheco-Vazquez2011} (exponent 2.38) and from us (exponent 2.45). It is worth to mention that crater volume-crater diameter plots are used in order to stress out the dimensionality of the physical excavation processes, which scales with a dynamic quantity (mass). It means that in these plots any restriction (amplification) in the excavation mechanism along any given direction should appear as a reduction (increasing) of the slope of the curve. Strikingly, the power law in the crater aspect ratio plots for terrestrial experiments performed during the last four decades with solid impactors or explosions does not obey the same power law of the observational data, which is more realistically reproduced by granular-granular collisions (Fig.~\ref{Exper}).

\begin{figure}[tb]
	\noindent\resizebox{\columnwidth}{!}{\includegraphics{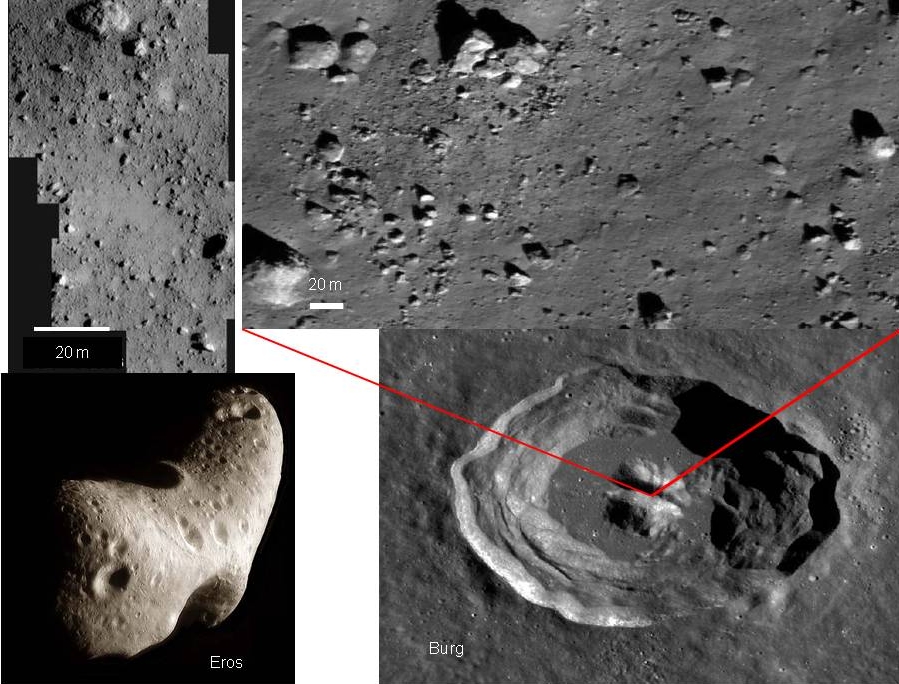}}
	\caption{\label{ErosBurg}Left images correspond to asteroid Eros. Right images correspond to crater Burg in the Moon. Note the similar granularities.}
\end{figure}

\begin{figure}[tb]
	\noindent\resizebox{\columnwidth}{!}{\includegraphics{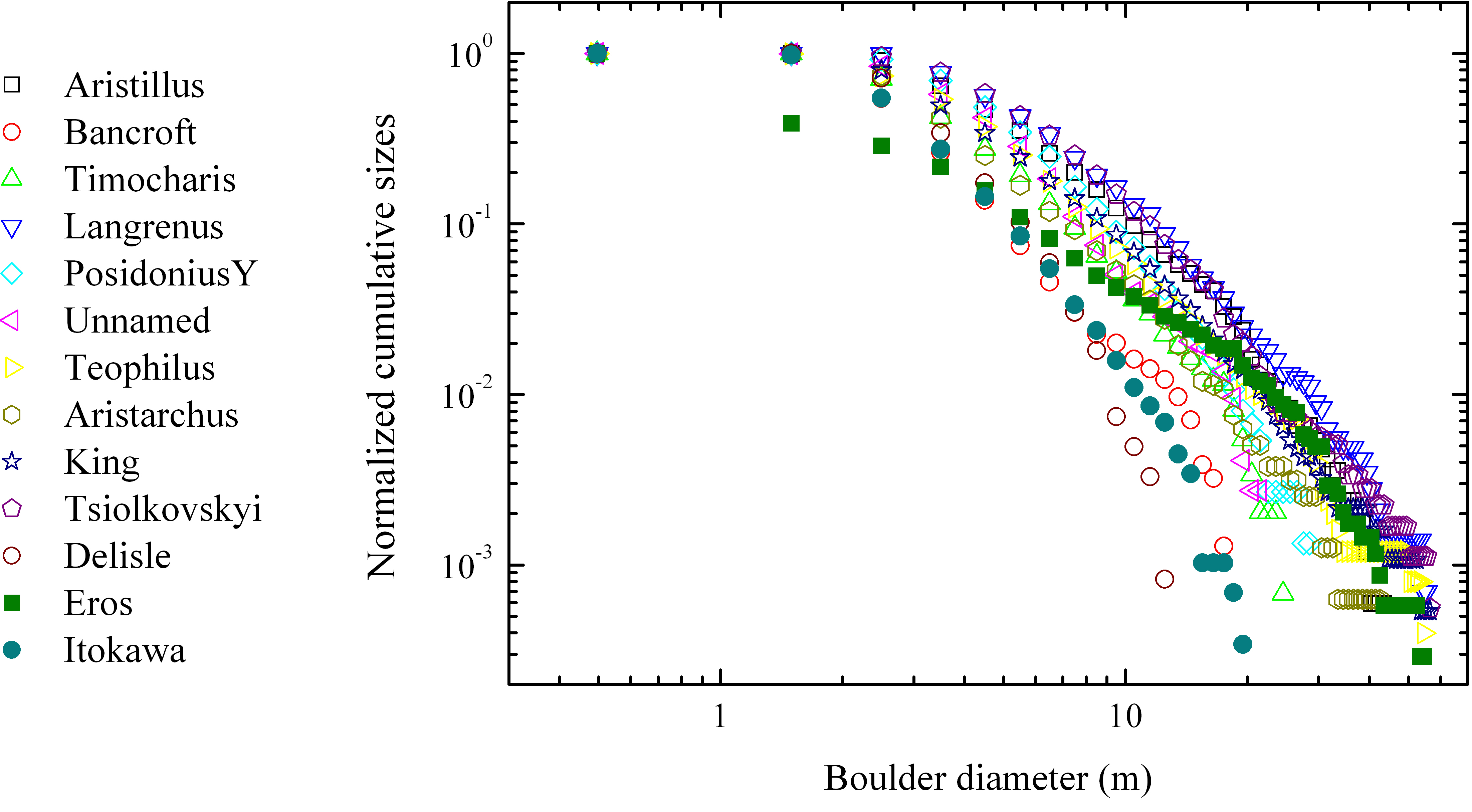}}
	\caption{\label{distr}Cumulative size distributions of rocks and boulders normalized relative to the total counted particles. For asteroids: Itokawa (from Saito \emph{et al.} \cite{Saito2006}) in solid circles and Eros in solid squares (obtained by us). For craters: hollow symbols. Coordinates of unnamed crater (magenta triangles) are latitude: $1.00672$, longitude: $-124.42014$}
\end{figure}

A photointerpretative analysis of high resolution images (50~cm/pixel) from the LROC probe of three central peak structures reveals the presence of large boulders (5--40~m) close to the top of the peak. In some cases such boulders roll down, showing tracks of their displacement, and evidencing a much smaller granulometry of the terrain. However, not a single signature of melting of the terrain or volcanic domes that could take such boulders to the top is observed. Analyzed at the light of our findings, we believe that such boulders were not moved upwards to the top by geological processes (as some authors have proposed \cite{Chauhan2012}), but instead, they werethey are fragments of the conminuted astroprotolith and deposited there by the dinamic confinement during the impact process. If the sequence of events during real impacts on the Moon, for example, is consistent with our hypotesis, we could see the evidence by analysing the central peaks and domes of the craters and also in the center of the floor of simple craters. As shown in fig 5, there are large boulders resting over a much (below limit resolution) finer material. In fig 6 we present the granulometric distribution of boulders that belong to craters with diameters from less than 2 km up to 200 km. If the impact process were due to melting and evaporation, no such granulometric distributions would be observed. To support this idea, we made a comparative analysis of cumulative size distribution of boulders on asteroid Eros (using high resolution photographs from the descent phase, see for example Fig.~\ref{ErosBurg} \cite{NotaItokawa}), and published data of Itokawa \cite{Saito2006}. Thereafter, we contrast them against the size distributions of boulders found close to the top of the central peaks and in the bottom of several lunar craters. As can be noted in Fig.~\ref{distr}, the size distributions are very similar, in accordance to the hypothesis that central peaks are vestigial remains of granular impactors deposited in the way described above and not by the rebound and uplift of the terrain at the center of the crater as claimed by current interpretations.


From the different power laws dependence of the crater volume as a function of the crater diameter followed by granular-granular and solid-granular collisions or explosive excavation of craters and the observed power law followed for impact craters in three satellites, we conclude that astroprotoliths impacting planetary bodies should share the same mechanism of crater formation than our laboratory granular-granular impacts. This means that compaction of the target by the comminuted/moldered impactor and its subsequent deposit within the crater reduce the otherwise expected power law (exponent around 3) to a smaller exponent. Observational and experimental evidences of complex crater morphologies suggest that their formation can be better explained in the light of granular-granular collisions and support the idea that domes and central peaks are remnants of impacting astroprotoliths dynamically confined during the collision event, and not the result of the transient crater collapse and uplift of the impacted terrain.

Compressional or dynamic pressure driven subsidence of the target and the deposition of the impactor debris, dynamically confined by granular superfast avalanches, sculpts complex crater morphologies observed in planetary bodies.

\section{Methodology}
\begin{figure}[tb]
	\noindent\resizebox{\columnwidth}{!}{\includegraphics{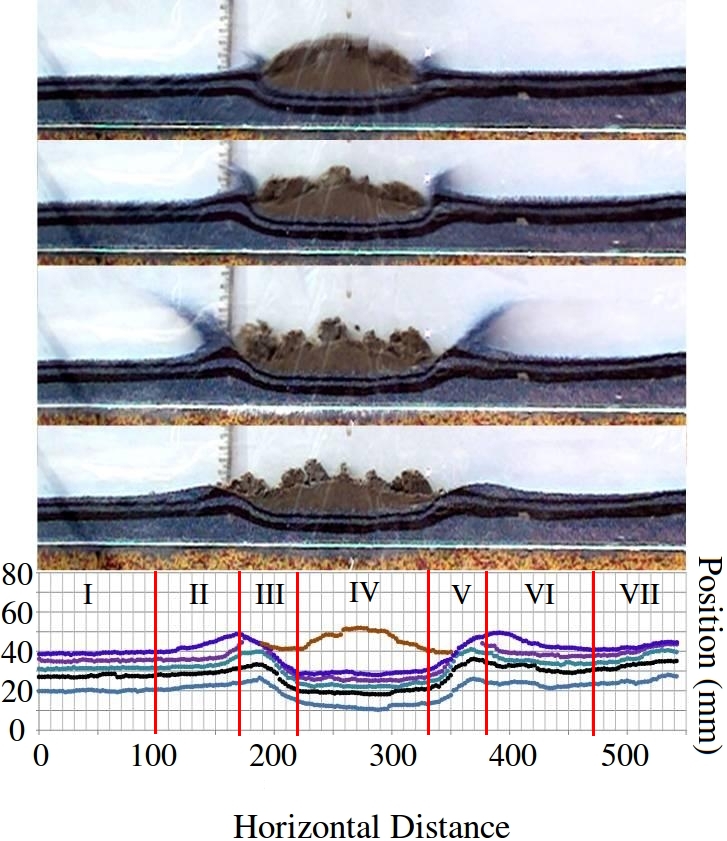}}
	\caption{\label{Sequence}Two-dimensional experiment of granular-granular crater formation as captured by a sequence of raw images. Bottom plot shows stratigraphy and regions of the crater studied: (I) left distal region, (II)left outer crater rim, (III) left inner crater rim, (IV) crater floor, (V) right inner crater rim, (VI) right outer crater rim and (VII) right distal region.}
\end{figure}
\subsection{3D experiments}
For 3D experiments the projectiles, with a range in diameters from 2 and 10~cm and packing fractions 0.24 and 0.27, were made from a slurry mixture composed of sand (grain size of  60--250~$\mu$m), polystyrene spheres (1~mm) and lead and stainless steel bearing balls (3~mm).  The main objective of introducing polystyrene spheres was to tailor controlling the compaction and porosity, since these beads shrink to half their size during the thermal treatment at 130~$^\circ$C of the projectiles. Lead and stainless steel spheres were introduced to increase the weight of the projectiles, in order to increase the kinetic energy of the impact. 

\begin{figure}[tb]
\begin{center}
	\noindent\resizebox{0.8\columnwidth}{!}{\includegraphics{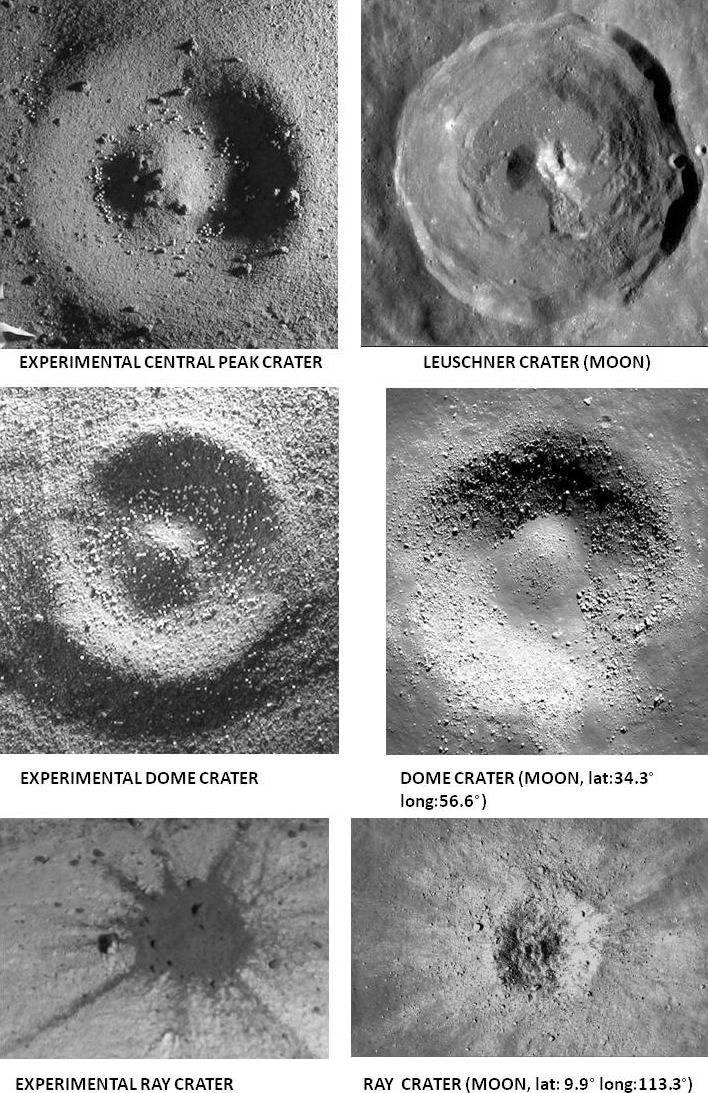}}
	\caption{\label{Craters}Left images correspond to craters observed in our experiments under different conditions. Right images correspond to craters on the Moon.}
\end{center}
\end{figure}

The experiments were performed dropping eight different sizes of our model astroprotoliths from a tower of 16~m total height, from different heights separated 3.5~m each. The resulting morphologies of the impact craters produced were measured using photographical techniques and their shapes are shown for selected craters in comparison with some similar craters observed in the Moon in Fig.~\ref{Craters}. There is a striking resemblance among them.

\subsection{2D experiments}
\begin{figure}[tb]
	\noindent\resizebox{\columnwidth}{!}{\includegraphics{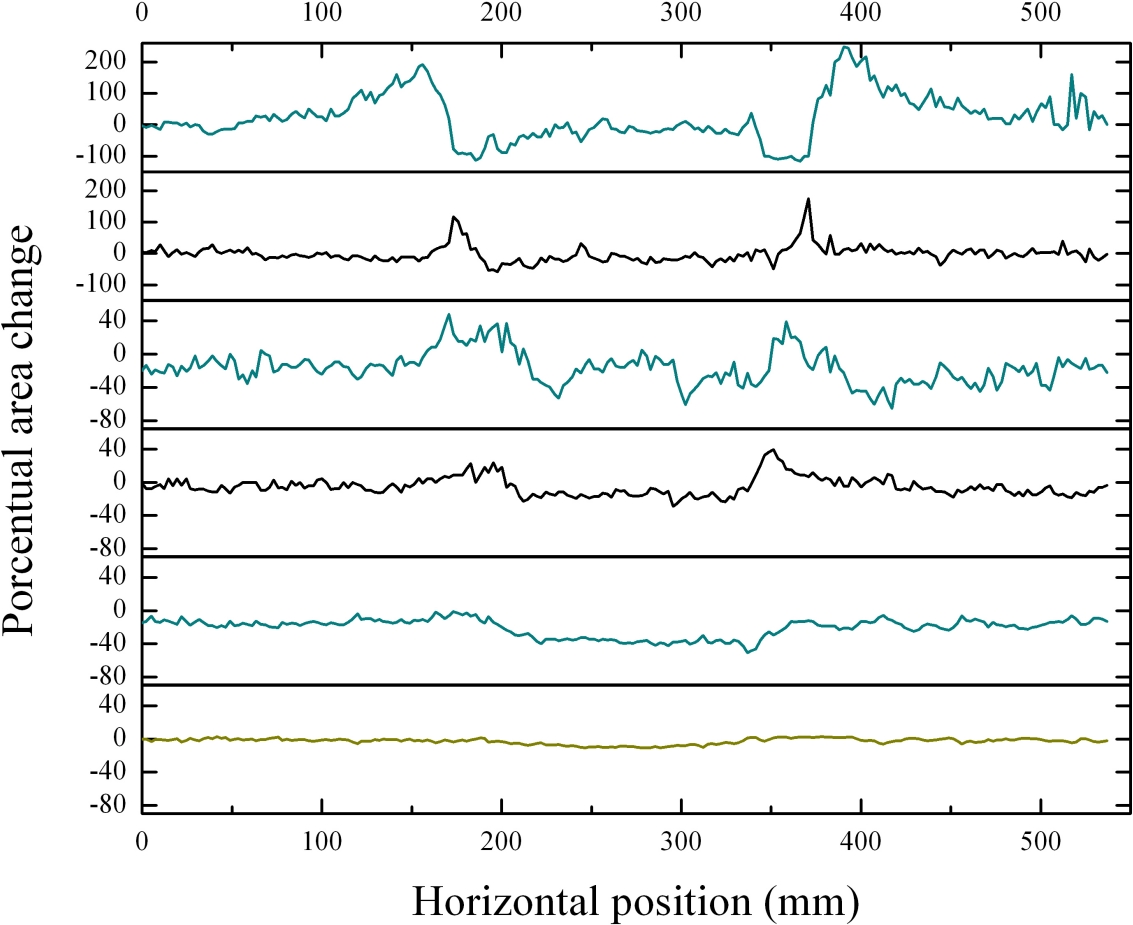}}
	\caption{\label{Strat}Area changes of subsequent strata as a function of their horizontal position measured after the collision event from a series of high speed digital videos. From top to bottom, the local area changes of their corresponding uppermost blue stratum, up to the deepest yellow stratum of the target in Fig.~\ref{Sequence}.}
\end{figure}

For 2D experiments, discs of the same mixture than above were prepared and dropped within a Hele-Shaw cell of 2~cm thick, 15~cm width and 7~m height. The target was a $89\times45\times2$~cm cell containing a color stratified bed of sand. Collisional events were recorded at 120 and 500~frames per second and analyzed by means of image analysis software.  The purpose of the strata was to analyze the compaction of the terrain due to the impact. This analysis was made by calculating the initial area of each stratum, and subsequently subtracting the area after the impact (Fig~\ref{Sequence} and \ref{Strat}). In a time close up, a superfast avalanche is observed during the formation of a central peak in some craters. 

The opening mechanism of a granular-against-granular collisional crater is depicted in Fig.~\ref{Diff} and Fig.~\ref{Sequence}, where a sequence of pictures of the crater formation process is presented. Images in Fig.~\ref{Diff} were subtracted by means of an Image J algorithm in order to stress out material mobilization during ejection or compression processes.

In order to visualize and characterize the opening crater process and its final morphology, we defined four symmetric regions of the crater: distal area; outer crater rim; left and right inner crater rim and crater floor. The thickness of each layer was measured on a frame before the impact and after the formation of the crater process was ended. We measured the full length of each layer with a spatial bin length of 2~mm.

At the crater floor and the inner and outer regions of the rim occurs the largest local changes in strata area due to three main reasons; material mobilization that will conform the ejecta, plastic deformation of the terrain and its consequent Reynolds dilatancy at the rim, and compaction of the terrain due to dynamic pressure exerted by the granular descending flux. Because of these three effects, percentage changes in area can be as high as 200\% (Fig.~\ref{Strat}).

The strata below the crater floor is highly compressed (subsidence) as can be better seen in Fig.~\ref{Diff}, in which differential photograms show a rectangular field of displacements that can be clearly noted just below the impactor at the first picture of the sequence. This displacement field broadens as time elapses until it affects the whole target in the cell.

The uppermost stratum at the inner crater rim region is completely ejected and deposited on top of the outer crater rim region. Deeper strata in this region is plastically deformed and thinned to conform the inner slope of the crater rim.

The uppermost stratum behaves the other way around at the outer crater rim region, since the ejecta is deposited on top of it. Meanwhile, deeper strata thicken due to Reynolds dilatancy. The distal region is slightly compressed for all strata, by the effect of the shock wave that rearranges the grains in the whole cell during the impact.

\begin{figure}[tb]
	\noindent\resizebox{\columnwidth}{!}{\includegraphics{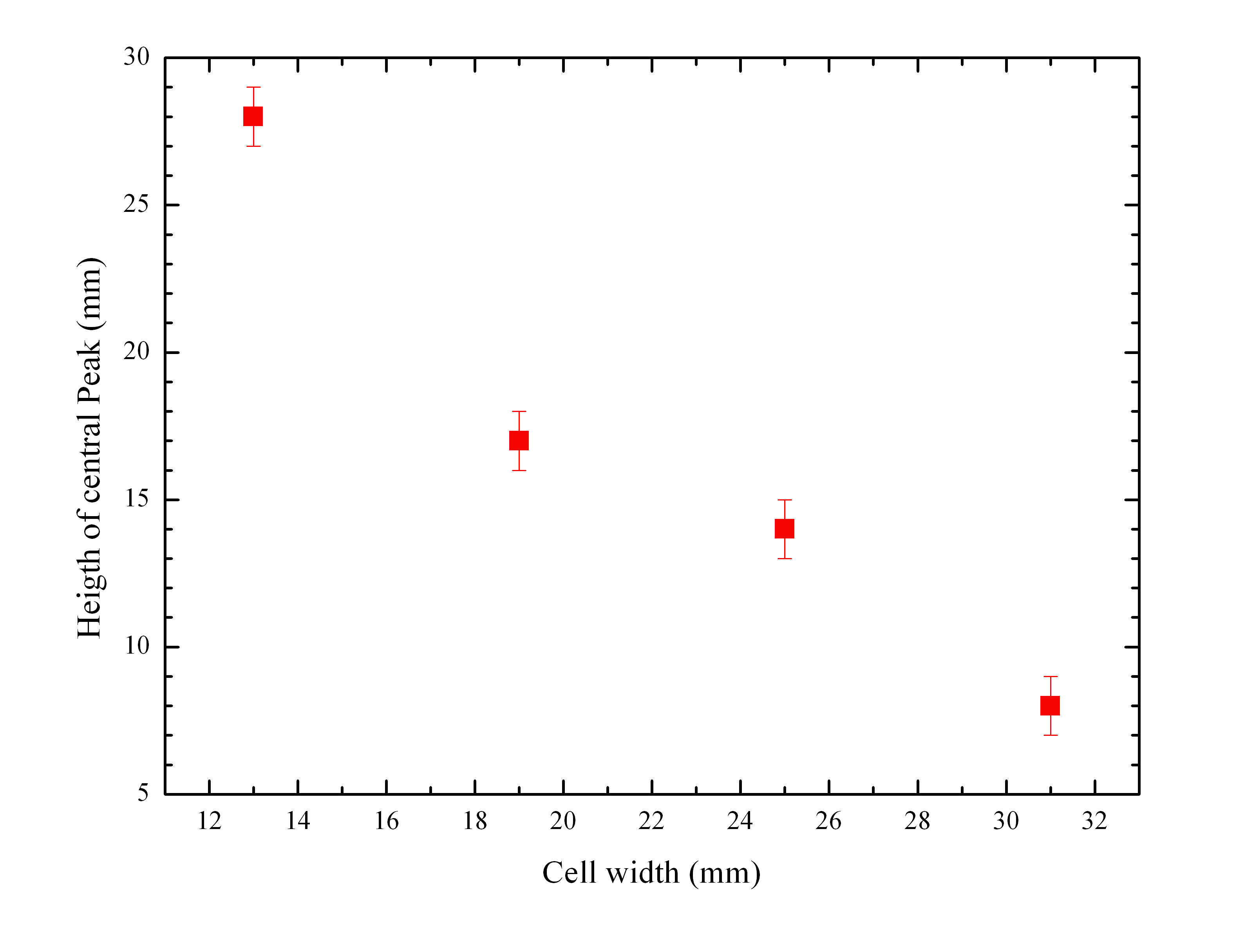}}
	\caption{\label{Height}Height of the central peak as a function of the width of the Hele-Shaw cell of the target, keeping constant the disc thickness, showing the reinforcement of the dynamic confinement by Janssen effect.}
\end{figure}

Since experiments were performed within a Hele-Shaw cell, the restriction imposed by the confining front and rear walls enhances the dynamic confinement due to Janssen's effect. This means that lateral confinement due to this cell configuration increases the height of confined material linearly with the cell width as shown in Fig.~\ref{Height}.

\subsection{Granulometric comparison between asteroids and central peaks}
Furthermore, an analysis of central peak structures of several craters reveals the presence of granulometric distributions consistent with the expected granulometry of astroprotoliths, as observed in the asteroid Eros. A visual comparison is shown in Fig.~\ref{ErosBurg}, which shows pictures of Eros and the central peak of the crater Burg at the Moon.

Boulder size measurements, both on the central peaks of lunar craters and on the surface of the asteroids Eros and Itokawa, were performed manually using Image Pro Plus software. We used several high resolution images of Eros from the database of NASA Near-Shoemaker mission (images Near Descent Mosaic 6288, 5228 and 6650) and in the case of Itokawa, we used images of the Japanese Hayabusa (133699128) mission. For the lunar crater we used high resolution images from the NASA LROC mission.

\begin{acknowledgments}
This work has been supported by Conacyt, Mexico, under Grants 82975, 46709 and 101384. R.B. and G.M.R.L. wish to acknowledge scholarships by CONACyT, Mexico.
\end{acknowledgments}

Author contributions: F.J.S.V., and J.C.R.S. designed research; F.J.S.V., and L.S.F.Q. performed research; F.J.S.V., L.S.F.Q., M.C., A.CH.R, and J.C.R.S. analyzed and discussed data; P.A.Z.M performed MD simulations; and F.J.S.V., M.C., and J.C.R.S. wrote the paper.

\bibliography{Crateres.bib}

\end{document}